\documentclass[12pt]{article}
\usepackage{amsfonts,latexsym}
\begin{document}
\newcommand{\de}{\delta}\newcommand{\ga}{\gamma}
\newcommand{\e}{\epsilon} \newcommand{\ot}{\otimes}
\newcommand{\be}{\begin{equation}} \newcommand{\ee}{\end{equation}}
\newcommand{\ba}{\begin{eqnarray}} \newcommand{\ea}{\end{eqnarray}}
\newcommand{\tmod}{{\cal T}}\newcommand{\amod}{{\cal A}}
\newcommand{\bemod}{{\cal B}}\newcommand{\cmod}{{\cal C}}
\newcommand{\dmod}{{\cal D}}\newcommand{\hmod}{{\cal H}}
\newcommand{\s}{\scriptstyle}\newcommand{\tr}{{\rm tr}}
\newcommand{\einsop}{{\bf 1}}
\def\oR{R^*} \def\upa{\uparrow}
\def\R{\overline{R}} \def\doa{\downarrow}
\def\oL{\overline{\Lambda}}
\def\nn{\nonumber} \def\dag{\dagger}
\def\beq{\begin{equation}}
\def\eeq{\end{equation}}
\def\bea{\begin{eqnarray}}
\def\eea{\end{eqnarray}}
\def\ve{\epsilon}
\def\si{\sigma}
\def\th{\theta}
\def\d{\delta}
\def\ga{\gamma}
\def\l{\left}
\def\r{\right}
\def\a{\alpha}
\def\b{\beta}
\def\g{\gamma}
\def\La{\Lambda}
\def\w{\overline{w}}
\def\u{\overline{u}}
\def\o{\overline}
\def\rr{\mathcal{R}}
\def\T{\mathcal{T}}
\def\N{\overline{N}}
\def\Q{\overline{Q}}
\def\L{\mathcal{L}}
\def\i{\overline{i}}
\def\j{\overline{j}}
\def\k{\overline{k}}
\def\l{\overline{l}}
\def\d{\dagger}
\newcommand{\reff}[1]{eq.~(\ref{#1})} 

\vspace{6cm}

\begin{center}
{\Large \bf 
Exact solution at integrable coupling
of a model for the Josephson effect between small metallic grains
}
\vskip.3in
{\large Jon Links$^*$, Huan-Qiang Zhou, Ross H. 
McKenzie$^+$ and Mark D. Gould}
\vskip.2in
{\em Centre for Mathematical Physics, \\
The University of Queensland,
      4072, \\ Australia

      $^*$email: jrl@maths.uq.edu.au
      
      $^+$Also at Centre for Quantum Computing Technology,
      The University of Queensland.}
      \end{center}

      \vskip 2cm
      \begin{abstract}
A model is introduced for two reduced BCS systems which are coupled
through the transfer of Cooper pairs between the systems. The model may
thus be used in the analysis of the Josephson effect arising from
pair tunneling between two strongly coupled small
metallic grains. At a particular coupling strength the model
is integrable and
explicit results are derived for the energy spectrum, conserved
operators, integrals of motion, and wave function scalar products.
It is also shown that form factors can be obtained for the
calculation of correlation functions.
Further, a connection with perturbed conformal field theory is made.
      \end{abstract}

\vfil\eject


\def\a{\alpha}
\def\b{\beta}
\def\d{\dagger}
\def\e{\epsilon}
\def\g{\gamma}
\def\k{\kappa}
\def\l{\lambda}
\def\o{\omega}
\def\t{\tilde{\tau}}
\def\s{S}
\def\D{\Delta}
\def\T{{\cal T}}
\def\TT{{\tilde{\cal T}}}

\def\beq{\begin{equation}}
\def\eeq{\end{equation}}
\def\bea{\begin{eqnarray}}
\def\eea{\end{eqnarray}}
\def\ba{\begin{array}}
\def\ea{\end{array}}
\def\no{\nonumber}
\def\le{\langle}
\def\re{\rangle}
\def\lt{\left}
\def\rt{\right}
\def\oR{R^*} \def\upa{\uparrow}
\def\R{\overline{R}} \def\doa{\downarrow}
\def\oL{\overline{\Lambda}}
\def\nn{\nonumber} \def\dag{\dagger}
\def\e{\epsilon}
\def\si{\sigma}
\def\th{\theta}
\def\de{\delta}
\def\ga{\gamma}
\def\l{\left}
\def\r{\right}
\def\a{\alpha}
\def\b{\beta}
\def\g{\gamma}
\def\La{\Lambda}
\def\w{\overline{w}}
\def\u{\overline{u}}
\def\o{\overline}
\def\rr{\mathcal{R}}
\def\T{\mathcal{T}}
\def\N{\overline{N}}
\def\Q{\overline{Q}}
\def\L{\mathcal{L}}
\def\i{\overline{i}}
\def\j{\overline{j}}
\def\k{\overline{k}}
\def\l{\overline{l}}
\def\d{\dagger}

\vspace{1cm}
    \centerline{\bf 1. Introduction.}
    \vspace{1cm}
    
The study of superconductivity in metallic nanoparticles
is of fundamental scientific interest and is the
basis of several proposed solid state quantum
computers \cite{schon}.
The experimental work of Ralph, Black and Tinkham
\cite{rbt} has stimulated a wealth of
theoretical activity concerning
the reduced BCS Hamiltonian describing
electron pairing correlations in systems with
discrete energy levels and a fixed number  of electrons.
For sufficiently small systems the energy
level spacing becomes comparable to the energy
gap predicted by the  BCS mean-field
approximation making mean-field theory unreliable because of the large
superconducting fluctuations \cite{dr01}.

These considerations are also relevant to
proposals to observe BCS superconductivity
in gases of fermionic atoms such as spin-polarised
$^6$Li \cite{stoof}. Quantum degeneracy of $^6$Li at
temperatures of about 240 nK has recently
been observed in an atom trap with frequencies,
$\omega \sim $ 1 kHz \cite{truscott}, corresponding
to an energy level spacing of the order of
$10^{-12}$ eV.
The estimated BCS transition temperature is
of the order of 20 nK \cite{stoof}, corresponding
to an energy gap of the order of 
$4 \times 10^{-12}$ eV, comparable to the energy
level spacing.

The failure of approximation schemes increases
the importance of the fact
that the reduced BCS Hamiltonian
is exactly solvable \cite{r63}.
Furthermore, it is integrable \cite{crs97},
 and has a rich
mathematical structure \cite{ao01,zlmg,vp}. 
The integrability of the model
has recently been clarified 
by the Quantum Inverse Scattering Method
(QISM) and algebraic Bethe ansatz \cite{zlmg,vp}. 
This has the advantage that it makes accessible the
computation of form factors and correlation functions \cite{ao01,zlmg}. 
This realisation makes it clear that these techniques can be used to study
more general classes of pairing models. For example, the coupled
pairing model which accomodates proton-neutron pairing in nuclear
systems studied by
Richardson \cite{r} can be formulated via the QISM utilizing the $so(5)$
symmetry algebra \cite{lzgm}. Other examples are discussed in
\cite{alo01,des}.

For a single metallic grain,
due to the fixed and finite number of electrons
there is no breaking of gauge symmetry and
no long-range superconducting order.
Consequently, an important question concerns the
existence and nature of the Josephson effect \cite{tinkham}
between two coupled metallic grains.
If the number of pairs of electrons $N$ in each grain
 is fixed, then for very weak intergrain
coupling there is no Josephson effect \cite{anderson}.
On the other hand, if one constructs a qubit
consisting of a coherent
superposition of $N$ and $N+1$ pairs in an individual
grain,
Josephson tunneling occurs and can be used to
couple qubits together \cite{schon,anderson}.

In this article we introduce a model describing
the tunneling of Cooper pairs between two
metallic grains described 
by reduced BCS models.
We show that for a particular value of the
intergrain coupling the model is integrable
and has an underlying so(3) symmetry which
we exploit to solve the model by the
algebraic Bethe ansatz.
First we consider the physical basis of the model,
stressing that it requires a large intergrain coupling.


\vspace{1cm}
    \centerline{\bf 2. Model Hamiltonian.}
    \vspace{1cm}
    
The physical properties of each metallic grain are described by the
reduced BCS Hamiltonian \cite{dr01}   
$$H_{BCS}(q)=\sum_{j=1}^{\L}\e_jn_j(q)-g\sum_{j,k}^{\L}b_k^{\d}(q)b_j(q) $$
with $q=1,\,2$.
Above, $j=1,...,{\L}$ labels a shell of doubly degenerate single particle
energy levels with energies $\e_j$ and $n_j$ is the 
fermion number operator for
level $j$. The operators $b_j\equiv 
  c_{j,\uparrow} c_{j,\downarrow}$
and $b_j^{\d}$ are the annihilation and
creation operators for the Cooper pairs and satisfy the hard-core boson
relations
$$(b^{\d}_j)^2=0, ~~[b_j,\,b_k^{\d}]=\delta_{jk}(1-2b^{\d}_jb_j) $$ 
$$[b_j,\,b_k]=[b^{\d}_j,\,b^{\d}_k]=0, ~~~{\rm for }~~ k\neq j. $$ 
One of the prominent features of this Hamiltonian is the {\it blocking
effect}. For any unpaired electron at level $j$ the action of 
the pairing interaction is zero since only paired electrons are
scattered. This means that the Hilbert space can be decoupled into 
orthogonal subspaces  of paired and unpaired electrons in which the
action of the Hamiltonian on the subspace for the unpaired electrons is 
automatically diagonal in the natural basis. Consequently, we need only
focus on  the subspace of paired electrons when solving for the spectrum
of the model.

In order to firmly establish a microscopic basis for
Josephson's proposal \cite{j,history} of Cooper pair tunneling
through an insulating barrier between two
bulk superconductors, 
Anderson \cite{anderson}, 
and Ambegaokar and Baratoff \cite{ab}, 
studied the Hamiltonian
\bea  
H&=&H_{BCS}(1)+H_{BCS}(2) 
 -  \sum_{j,k,\sigma} t_{jk}\left(c^{\d}_{j,\sigma}(1)c_{k,\sigma}(2)+
c^{\d}_{j,\sigma}(2)c_{k,\sigma}(1)\right). \label{tham} \eea    
They treated the effect of the tunneling to
second-order in $t_{jk}$ and the BCS Hamiltonians
were treated at the mean-field level in the
grand-canonical ensemble.
If we define $\Delta \exp(i \phi(q)) \equiv g \sum_j
 \langle c^{\d}_{j,\uparrow}(q)
c^{\d}_{j,\downarrow}(q) \rangle$
then the Josephson current is
proportional to 
$ t^2 \Delta \sin (\phi(1) - \phi(2))$
where $t^2 \equiv \sum_{j,k} |t_{jk}|^2$.
One is then justified in writing the effective Hamiltonian 
\bea  
H&=&H_{BCS}(1)+H_{BCS}(2) 
 - \varepsilon_J \sum_{j,k}^{\L}\left(b^{\d}_j(1)b_k(2)+
b^{\d}_j(2)b_k(1)\right) \label{jham} \eea    
 where $\varepsilon_J$ is the Josephson coupling energy.

If one repeats the calculation with $N(1)$ and $N(2)$ fixed
and finite
then to second-order in the tunneling matrix element
there is no Josephson coupling because
 $\langle c^{\d}_{j,\uparrow}(q)
c^{\d}_{j,\downarrow}(q) \rangle = 0$
for both $q=1, 2$.
However, if the intergrain coupling is sufficiently
large that we go beyond second-order in $t$ then only
$N(1) + N(2)$ is a good quantum number, and not
$N(1)$ and $N(2)$ individually. It
is then possible that
 $\langle c^{\d}_{j,\uparrow}(q)
c^{\d}_{j,\downarrow}(q) \rangle \neq  0$
and so it is appropriate
to study a Hamiltonian of the form (\ref{jham}).


The Hamiltonian (\ref{jham}) is not integrable for arbitrary values of
the coupling parameters $g$ and $\varepsilon_J$.
 We demonstrate below that for the
particular choice $\varepsilon_J=g$
 the model does become integrable and an exact
solution can be obtained.
In typical Josephson junctions involving
spatially separated superconductors
that are separated by an insulating barrier, $\varepsilon_J \ll g$.
However, this does not rule out
fabrication of a junction consisting
of two metallic grains separated by
a quantum point contact with a transmission coefficient
close to one.
Because of the tunability of the pairing
interaction in trapped atom systems, as noted in \cite{truscott}, the case of 
(relatively) strong Josephson coupling between such systems 
may be feasible.
These systems also admit the possibility of
an ``internal'' Josephson junction which
does not involve spatially separated condensates \cite{leggett}.

\vspace{1cm}
    \centerline{\bf 3. Integrable coupling model.}
    \vspace{1cm}

As in the case of a single BCS system, the model (\ref{jham}) 
exhibits the blocking
effect for unpaired electrons. However, the model with $\varepsilon_J=g$ 
displays an additional
blocking effect which we now elucidate.
 We begin by writing the Hamiltonian
in an equivalent form 
\beq H=\sum_{j=1}^{\L}\e_jn_j-g\sum^{\L}_{j,k}(b^{\d}_j(1)+b^{\d}_j(2))
(b_k(1)+b_k(2)) \label{ham} \eeq  
where $n_j=n_j(1)+n_j(2)$.  
The pairing interaction cannot scatter any   
antisymmetric state of the form   
$$\left|\Psi_j\right>= 
\frac{1}{\sqrt 2}\left(b^{\d}_j(1)- b^{\d}_j(2)\right)\left|0\right>$$ 
since an elementary calculation shows that 
$$(b_k(1)+b_k(2))\left|\Psi_j\right>=0 $$ 
for all $k$ and $j$.
A consequence of the blocking effect for antisymmetric pairs is that the
Hilbert space now decomposes into subspaces where the pairing
interaction is non-trivial on the local states spanned by 
\bea \left|1\right>&=&b^{\d}(1)b^{\d}(2)\left|0\right>, \nn \\
\left|2\right>&=&\frac{1}{\sqrt 2}\left(b^{\d}(1)+ 
b^{\d}(2)\right)\left|0\right>, \nn \\
\left|3\right>&=&\left|0\right>  \nn \eea 
for each level $j$. We will denote this space $V_j$. 
Hence the problem of diagonalizing the Hamiltonian is reduced to
diagonalizing the Hamiltonian over the product of 
these subspaces. This will be achieved
below using the algebraic Bethe ansatz and the spin 1 representation of
the $so(3)$ Lie algebra. The procedure is analogous to that employed in
Ref. \cite{zlmg,vp} for the standard reduced BCS model
involving $su(2)$.

The hard-core bosons provide a spin 1 representation   
of $so(3)$ through 
\bea   
S^+= \left(b^{\d}(1)+b^{\d}(2)\right), ~~  
S^-= \left(b(1)+b(2)\right), ~~ 
S^z&=&N-1 \label{rep}  \eea        
where $N=N(1)+N(2)$ counts the number of Cooper pairs. 
Associated with the spin 1 representation there is a solution 
of the Yang-Baxter equation 
$$R_{12}(u-v)L_{1}(u)L_{2}(v)=L_{2}(v)L_{1}(u)R_{12}(u-v)$$ 
with 
\bea 
R(u)&=&I\otimes I +\frac{\eta}{u}\sum_{m,n=1}^2 e^m_n \otimes e^n_m 
, \nn \\
L(u)&=&I\otimes I + \frac{\eta}{u}\left(e^1_1\otimes S^z
-e^2_2\otimes S^z 
+e^1_2\otimes S^-+e_1^2\otimes S^+ \right) \nn 
\eea 
where $\{e^m_n\}$ are $2\times 2$ matrices with 1 in the
$(m,n)$ entry and zeroes eleswhere. 
Above, $I$ is the identity operator and 
$\eta$ is a scaling parameter for the rapidity variable $u$ which
plays an important role in the subsequent analysis. 
With this solution we construct the transfer matrix 
\beq t(u)={\rm tr}_0\left(G_0L_{0{\L}}(u-\e_{\L})...L_{01}(u-\e_1)\right) 
\label{tm} \eeq  
which is an operator on the subspace $V_1\otimes V_2\otimes ... V_{\L}$ on
which the pairing interaction acts non-trivially. Above, we have
introduced an auxiliary space $V_0$ and ${\rm tr}_0$ denotes the trace
taken over this space. Moreover $G=\exp(\alpha\eta {\sigma})$ with
$\sigma={\rm diag}(1,\,-1)$. 
A consequence of the Yang-Baxter equation is that 
$[t(u), \,t(v)]=0$ 
for all values of the parameters $u$ and $v$. 
Defining 
$$T_j=\lim_{u\rightarrow \e_j}\frac{u-\e_j}{\eta^2} t(u) $$ 
for $j=1,2,...,{\L}$,
we may write in the {\it quasi-classical limit}  
$T_j=\tau_j+o(\eta) $ 
and it follows that 
$$ [\tau_j,\,\tau_k]=0, ~~~~ \forall j,\,k. $$ 
Explicitly, these operators read 
\beq 
\tau_j=2\alpha S^z_j+\sum_{k\neq j}^{\L}\frac{\theta_{jk}}{\e_j-\e_k}
\label{cons} \eeq 
with 
$\theta=S^+\otimes S^-+S^-\otimes S^++2S^z\otimes S^z.$ 
It is an algebraic exercise to show that the Hamiltonian (\ref{ham}) 
can be expressed as 
\bea  
H&=&-g\sum_j^{\L}\e_j\tau_j+g^3/4\sum_{j,k}^{\L}\tau_j\tau_k 
 +2\sum_j^{\L}\e_j
+g^2/2\sum_j^{\L} \tau_j- 2g{\L} 
\label{ham1} \eea       
with $g=-1/\alpha$ and the elements of $so(3)$ represented through
(\ref{rep}).  
This shows that the Hamiltonian is integrable since it is clear that 
$[H,\,\tau_j]=0,~~\forall j. $  

\vspace{1cm}
    \centerline{\bf 4. Exact solution.}
	\vspace{1cm}
	
A standard calculation
(e.g., see \cite{fad}) gives that the eigenvalues of the transfer matrix
(\ref{tm}) take the form
\bea
\Lambda(u)&=&\exp(\alpha\eta)\prod_k^{\L}\frac{u-\e_k+\eta}{u-\e_k}
\prod_j^M\frac{u-w_j-\eta}{u-w_j} \nn \\
&&~
+\exp(-\alpha\eta)\prod_k^{\L}\frac{u-\e_k-\eta}{u-\e_k}
\prod_j^M\frac{u-w_j+\eta}{u-w_j} \nn  
 \eea
where the parameters $w_j$ are required to satisfy the Bethe ansatz
equations
$$\exp(2\alpha\eta)\prod_k^{\L}\frac{w_l-\e_k+\eta}{w_l-\e_k-\eta}
=-\prod_j^M\frac{w_l-w_j+\eta}{w_l-w_j-\eta} . $$

The eigenvalues of the conserved operators (\ref{cons}) are obtained
through the leading terms of the expansion of the transfer matrix eigenvalues  
in the parameter $\eta$. This yields the result 
$$\lambda_j= \frac{-2}{g} +\sum_{k\neq j}^{\L}\frac{2}{\e_j-\e_k} -\sum_i^M 
\frac{2}{\e_j-v_i} $$ 
such  that the parameters $v_j$ satisfy the coupled algebraic equations 
\beq 
\frac{-2}{g}+ \sum_k^{\L}\frac{2}{v_j-\e_k}=\sum_{i\neq j}^M \frac{2}{v_j-v_i}
\label{bae} \eeq  
which are analogous to Richardson's equations \cite{r63} 
for the reduced BCS model.
Through (\ref{ham1}), the energy eigenvalues are found to be 
$$E=4\sum_j^{\L} \e_j- 2\sum_j^M v_j +2g(M-{\L}). $$ 

We introduce a set of states of the form 
\beq \left|\vec{w},M\right>
\equiv \left|w_1,...w_M\right>=\prod_j^M B(w_j)\left|\Phi\right> 
\label{states} \eeq  
where $\left|\Phi\right>$ is the completely filled state with $2{\L}$ Cooper
pairs and
$$B(u)=\sum_j^{\L}\frac{S^-_j}{u-\e_j}.$$
Note that these operators are mutually commuting for different values of
$u$. 
In general we find that 
\beq
\tau_j \left|\vec{w},M\right> = \lambda_j \left|\vec{w},M\right> -
\sum ^M_{\a} \frac {f_\a \s^-_j}{\e_j-w_\a}
\left|\vec{w},M\right>'_\a, \label{qc-osbae}
\eeq
where
\beq
f_\a = \frac {2}{g} +  \sum^M_{\b \neq \a} \frac {2}{w_\a -w_\b}
- \sum^{\L}_j \frac {2}{w_\a -\e_j} \no
\eeq
and $\lambda_j$ is understood to be a function of the parameters
$w_i$.
In (\ref{qc-osbae}) we defined
$\left|\vec{w},M\right> '_\a$ by
$$\left|\vec{w},M\right> = \sum^{\L} 
_{j} \frac {\s ^-_j}{w_\a-\e_j} \left|\vec{w},M\right>'_\a. $$
Imposing $f_\a =0$, one immediately sees that $\left|\vec{w},M\right>$ 
becomes the
eigenvector of the integrals of motion $\tau _j$ with
$\lambda_j$ as the eigenvalue. The constraint $f_\a=0$ is then
equivalent to (\ref{bae}).  
For a given solution of (\ref{bae}), the corresponding eigenstate has
$2{\L}-M$ Cooper pairs.

\vspace{1cm}
    \centerline{\bf 5. Connection to conformal field theory.}
	\vspace{1cm}

Introduce the function $\chi (v,\e)= \chi (v_1,\cdots,v_N,
\e_1,\cdots,\e_\Omega)$ obeying the following differential relations
\beq 
\kappa \frac {d \chi(v,\e)}{d\e_j} = \lambda_j \chi(v,\e),~~~~ 
\kappa \frac {d \chi(v,\e)}{dv_\a} = f_\a \chi(v,\e), \label{de}
\eeq 
where $\kappa = (k+2)/2$. It is easy to check that the zero curvature
conditions are fulfilled: 
$$\frac {d\lambda_j}{dv_\a}=\frac {df_\a}{d\e_j}.$$
The solution of (\ref{de}) is
\bea 
\chi (v,\e)& =& \exp\left(-\frac {2}{g \kappa} \sum _j \e_j + 
\frac {2}{g \kappa} \sum _\a v_\a\right) \prod _{i < j} (\e_i -\e_j) ^{\frac
{2}{\kappa}} \no \\
&&~\times \prod _{\a < \b} (v_\a -v_\b)^{\frac {2}{\kappa}}
\prod _{j\a} (\e_j - v_\a)^{- \frac {2}{\kappa}}.
\no \eea 
The function defined through multiple contour integrals 
\beq
\Phi (\e_1, \cdots, \e_\Omega) = \oint \cdots \oint
\chi (v,\e) \psi (v,\e) dv_1 \cdots dv_N
\eeq
generates a solution to the (perturbed) Knizhnik-Zamolodchikov
(KZ) equation \cite{kz84}
$$
\kappa \frac {d \Phi}{d\e_j} =  \tau _j \Phi.
$$
This constitutes a generalization of the solutions of the KZ equation
found in \cite{b93} (see also, \cite{cf87,djmm90,sv91}).
It should be stressed that the perturbed KZ equation is understood in
the sense of perturbed conformal field theory \cite{za}. That is,
the conformal invariance is broken once the term $-2S^z_j/g$
is included. The above construction is an extension of the corresponding
results for the BCS model, which is equivalent to Sierra's free field 
realization of the $su(2)_k$ WZNW model \cite{s00}. As Sierra noted for the
BCS model,
the on-shell eigenvalue of the integrals of motion of the model
follows from the saddle point approximation in the singular limit
$k=-2$.

\vspace{1cm}
    \centerline{\bf 6. Scalar products and norms.}
	\vspace{1cm}
	
There is an elegant result due to Slavnov 
\cite{sla89,kib,kmt98} which gives a determinant representation for the
scalar product of an eigenstate of the transfer matrix (\ref{tm}) with an
arbitrary state. 
In the quasiclassical limit, the leading
terms in $\eta$ of Slavnov's scalar product  
gives rise to the scalar product of the states of this model yielding
the result 
\bea
&& \langle \vec{w},M|\vec{v},M\rangle = 
\frac {\prod ^M_{\b=1} \prod ^M_{\stackrel {\a=1}{\a \neq \b}}
(v_\b-w_\a)}
{\prod _{\b <\a} (w_\b -w_\a) \prod _{\a <\b} (v_\b -v_\a)}
{\rm det}_M J(\{ v_\a \}, \{ w_\b \}),\nn 
\eea
where the matrix elements of $J$ are given by
\bea
J_{ab} &=& \frac {v_b -w_b}{v_a -w_b} 
\left ( \sum ^{\L} _{j=1} \frac {2}
{ (v_a -\e _j)(w_b -\e _j)} \right. \no\\
&& \left. -\sum^M _{\a \neq a} \frac {2}{(v_a
-v_\a)(w_b -v_\a)} \right ).
\label{sp}
\eea
Here $\{ v_\a \}$ are a solution to the equations (\ref{bae}),
whereas $\{ w_\b \}$ are arbitrary parameters.
Specializing $\{w_b\}=\{v_a\}$ for a solution to (\ref{bae}) 
gives the square of the norm of the
corresponding eigenvector (\ref{states}),
\beq
 \langle \vec{v},M|\vec{v},M\rangle = 
{\rm det}_M K,\nn 
\eeq
where the matrix elements of $K$ are given by
\bea
K_{aa} &=&  
\sum ^{\L} _{j=1} \frac {2}
{ (v_a -\e _j)^2} 
-\sum^M _{\a \neq a} \frac {2}{(v_a-v_\a)^2}, \no\\
K_{ab} &=&  
\frac {2}{(v_a -v_b)^2}. 
\eea

\vspace{1cm}
    \centerline{\bf 7. Correlation functions and form factors.}
	\vspace{1cm}
	
For any operator $\chi$ and state $\left|\vec{w},M\right>$
the correlation function is defined
$$C(\chi, \vec{w},M)=\frac{\left<\vec{w},M|\chi|\vec{w},M\right>}
{\left<\vec{w},M|\vec{w},M\right>} $$
Of
particular interest in the present case are the correlation functions
for the intergrain current operator
\bea 
\bf{j}&=&i[H,\,\sum_j n_j(1)-n_j(2)] \nn \\ 
&=&2ig\sum_{j,k}\left(b^{\d}_j(1)b_k(2)-b^{\d}_j(2)b_k(1)\right)\nn \\
&=&ig\sum_{j,k}\left(S^+_jW^-_k-W^+_jS^-_k\right) \nn \eea 
where $W^+=b^{\d}(1)-b^{\d}(2),\,W^-=b(1)-b(2).$ In the case when
$\left|\vec{w},M\right>$ is an eigenvector of the
Hamiltonian it can be demonstrated that $C({\bf{j}},\vec{w},M)=0$. For
general
states however it is non-zero. In order to evaluate the expectation
value of the current for a general state requires knowing the off-diagonal
form factors.

We can use (\ref{sp}) to obtain expressions for 
off-diagonal form
factors for the spin operators. Observe that the inverse problem for the
operators $S^{-}_m$
is easily solved in the present case through
\beq
S^-_m=\lim_{u\rightarrow \e_m}(u-\e_m)B(u). \label{inv} \eeq
A consequence is that the following off-diagonal form factors
$$ \left<\vec{w},M|S^+_m|\vec{v},M+1\right> 
= \left<\vec{v},M+1|S^-_m|\vec{w},M\right>
$$
can be computed directly from (\ref{sp}).
We find the following expression     
\bea
&&\left<\vec{w},M+1|S^-_m|\vec{v},M\right> 
 =\frac {\prod ^{M+1}_{\b=1} (w_\b - \e_m)} 
{\prod ^M_{\a=1} (v_\a - \e_m)} 
\frac { {\rm det}_{M+1} {\T} (m, \{ w_\b \}, \{ v_\a \})}
{\prod _{\b > \a} (w_\b -w_\a) \prod _{\b <\a} (v_\b -v_\a)}.\label{ff}
\eea
with the matrix elements of $\T$ given by
\bea
{\T}_{ab}(m) =&& 
\prod ^{M+1}_{\stackrel {\a=1}{\a \neq a}} (w_\a - v_b)
\left ( \sum ^{\L} _{j=1} \frac {2}
{ (v_b -\e _j)(w_a -\e _j)} \right. \no\\ 
&& \left. -\sum^M _{\a \neq a} \frac {2}{(v_b-
w_\a)(w_a -w_\a)} \right ),
~~~b < M+1, \no\\
{\T}_{a,M+1}(m)  && =  \frac {2}{(w_a -\e_m)^2}, \ \
\no \eea
The problem of obtaining analogous expressions for the operators 
$W^+_m,\,W^-_m$ is not as straightforward and we will return to this
problem at a later date.
We note that general correlation functions and form factors can be
computed using
the
leading terms in $\eta$ (quasi-classical limit) of the results 
derived by  Kitanine \cite{kit} for the spin
1 Takhtajan-Babujian model. 

\vspace{1cm}
    \centerline{\bf 8. Conclusion.}
	\vspace{1cm}

We have proposed a model describing strong Josephson tunneling in
coupled small metallic grains 
which are described by reduced BCS Hamiltonians. 
At a particular value of the Josephson coupling
energy we have demonstrated that the model becomes integrable, which
allows for the exact calculation  of the energy spectrum and form
factors of the model. We believe that these results will provide insight
into the nature of the Josephson effect at the nanoscale level where
mean field approaches are not applicable.

\vskip.3in
\begin{flushleft}
{\bf Acknowledgements} - This work was supported by the Australian Research Council.
\end{flushleft}

\end{document}